\documentclass[twocolumn,preprintnumbers,amsmath,amssymb]{revtex4}

\usepackage{graphicx}% Include figure files
\usepackage{dcolumn}% Align table columns on decimal point
\usepackage{bm}% bold math
\usepackage{rotating}

\usepackage{bbm}
\usepackage{verbatim}
\usepackage{afterpage}
\usepackage{color}
\definecolor{lightblue}{rgb}{0.2,0.2,0.7}
\definecolor{darkblue}{rgb}{0,0.25,0.5}
\definecolor{redbrown}{rgb}{0.875,0.25,0.125}
\definecolor{darkgreen}{rgb}{0,0.5,0}

\newcommand{\bra}[1]{\ensuremath{\langle #1 \vert}}
\newcommand{\ket}[1]{\ensuremath{\vert #1  \rangle}}

\renewcommand{\b}[1]{\ensuremath{\mathbf{#1}}}
\renewcommand{\H}{\ensuremath{\text{H}}}
\renewcommand{\l}{\ensuremath{\lambda}}

\newcommand{\HF}{\ensuremath{\text{HF}}}

\newcommand{\g}{\ensuremath{\gamma}}

\begin{document}

\title{Double-hybrid density-functional theory with meta-generalized-gradient approximations}

%%%%%%%%%%%%%%%%%%%%%%%%%%%%%%%%%%%%%%%%%%%%%%%%%%%%%%%%%%%%%%%%%%%%%%%%%%%%%
\author{Sidi Ould Souvi$^{1,2}$\footnote{Present address: Institut de Radioprotection et S\^uret\'e Nucl\'eaire, PSN-RES/SAG/LETR, Cadarache, 13115 Saint-Paul-l\`es-Durance, France}}\email{sidi.souvi@irsn.fr}
\author{Kamal Sharkas$^{1,2}$}
\author{Julien Toulouse$^{1,2}$}\email{julien.toulouse@upmc.fr}
\affiliation{
$^1$Sorbonne Universit\'es, UPMC Univ Paris 06, UMR 7616, Laboratoire de Chimie Th\'eorique, F-75005 Paris, France\\
$^2$CNRS, UMR 7616, Laboratoire de Chimie Th\'eorique, F-75005 Paris, France}

%%%%%%%%%%%%%%%%%%%%%%%%%%%%%%%%%%%%%%%%%%%%%%%%%%%%%%%%%%%%%%%%%%%%%%%%%%%%%

\date{\today}

\begin{abstract}
We extend the previously proposed one-parameter double-hybrid density-functional theory [K. Sharkas, J. Toulouse, and A. Savin, J. Chem. Phys. {\bf 134}, 064113 (2011)] to meta-generalized-gradient-approximation (meta-GGA) exchange-correlation density functionals. We construct several variants of one-parameter double-hybrid approximations using the Tao-Perdew-Staroverov-Scuseria (TPSS) meta-GGA functional and test them on test sets of atomization energies and reaction barrier heights. The most accurate variant uses the uniform coordinate scaling of the density and of the kinetic energy density in the correlation functional, and improves over both standard Kohn-Sham TPSS and second-order M{\o}ller-Plesset calculations.
\end{abstract}

\maketitle

\section{Introduction}

The double-hybrid (DH) approximations~\cite{Gri-JCP-06} have become ones of the most accurate approximations for electronic-structure calculations within density-functional theory (DFT). They consist in mixing Hartree-Fock (HF) exchange with a semilocal exchange density functional and second-order M{\o}ller-Plesset (MP2) correlation with a semilocal correlation density functional:
\begin{eqnarray}
E_{xc}^{\text{DH}} &=& a_x E_{x}^{\HF} + (1-a_x) E_{x}[n]
\nonumber\\
&& + (1-a_c) E_{c}[n] + a_c E_{c}^{\text{MP2}},
\label{DH}
\end{eqnarray}
where the first three terms are calculated in a self-consistent hybrid Kohn-Sham (KS) calculation, and the last MP2 term is usually evaluated with the previously obtained orbitals and added \textit{a posteriori} (see, however, Ref.~\onlinecite{PevHea-JCP-13} for a double-hybrid scheme with orbitals optimized in the presence of the MP2 correlation term). The two empirical parameters $a_x$ and $a_c$ are usually determined by fitting to a thermochemistry database. A variety of such double hybrids have been constructed with different $a_x$ and $a_c$ parameters and various density functionals~\cite{Gri-JCP-06,SchGri-PCCP-06,GraMenGoeGriRad-JPCA-09,TarKarSerVuzMar-JPCA-08,KarTarLamSchMar-JPCA-08,GraMenGoeGriRad-JPCA-09,GoeGri-JCTC-11,Yu-IJQC-13}. Double-hybrid approximations with more parameters have also been proposed~\cite{ZhaXuGod-PNAS-09,ZhaLuoXu-JCP-10,KozGruMar-JPCC-10,ZhaXuJunGod-PNAS-11,KozMar-PCCP-11,ZhaSuBreAdaXu-JCP-12,KozMar-JCC-13}. The so-called multicoefficient correlation methods combining HF, DFT and MP2 energies can also be considered to be a form of double-hybrid approximation~\cite{ZhaLynTru-JPCA-04,ZhaLynTru-PCCP-05,ZheZhaTru-JCTC-09,SanPer-JCP-09}. 

Recently, Sharkas {\it et al.}~\cite{ShaTouSav-JCP-11} provided a rigorous theoretical justification for double hybrids based on the adiabatic connection formalism and which lead to a density-scaled one-parameter double-hybrid (DS1DH) approximation
\begin{eqnarray}
E^{\text{DS1DH},\l}_{xc} &=& \l E_x^{\HF} + (1-\l) E_x[n] 
\nonumber\\
&&+ E_c[n] -\l^2 E_c[n_{1/\l}] + \l^2 E_c^{\text{MP2}},
\label{ExcDS1DH}
\end{eqnarray} 
where $E_c[n_{1/\l}]$ is the usual correlation energy functional evaluated at the scaled (squeezed) density $n_{1/\l}(\b{r})=(1/\l)^3 n(\b{r}/\l)$. In this class of double hybrids only one independent empirical parameter $\l$ is needed instead of the two parameters $a_x$ and $a_c$. The connection with the original double-hybrid approximations can be made by neglecting the density scaling
\begin{eqnarray}
E_c[n_{1/\l}] \approx E_c[n],
\label{Ecnoscaled}
\end{eqnarray} 
which leads to the one-parameter double-hybrid (1DH) approximation~\cite{ShaTouSav-JCP-11}
\begin{eqnarray}
E^{\text{1DH},\l}_{xc} &=& \l E_x^{\HF} + (1-\l) E_x[n] 
\nonumber\\
&&+ (1-\l^2) E_c[n] + \l^2 E_c^{\text{MP2}}.
\label{Exc1DH}
\end{eqnarray} 
corresponding to the standard double-hybrid approximation of Eq.~(\ref{DH}) with parameters $a_x=\l$ and $a_c=\l^2$. DS1DH and 1DH approximations have been constructed using the Perdew-Burke-Ernzerhof (PBE)~\cite{PerBurErn-PRL-96} and the Becke-Lee-Yang-Parr (BLYP)~\cite{Bec-PRA-88,LeeYanPar-PRB-88} exchange-correlation density functionals, and it was found that when neglecting density scaling the accuracy on atomization energies largely deteriorates for PBE but in fact improves for BLYP (see, also, Ref.~\onlinecite{Ali-JPCA-13} for a 1DH approximation based on the modified Perdew-Wang~\cite{AdaBar-JCP-98} exchange functional and the Perdew-Wang-91~\cite{Per-INC-91} correlation functional). By appealing to the high-density limit of the correlation functional, Toulouse {\it et al.}~\cite{TouShaBreAda-JCP-11} argued that a more sensible approximation to the density-scaled correlation functional is
\begin{eqnarray}
E_c[n_{1/\l}] \approx (1-\l) E_c^{\text{MP2}} + \l E_c[n],
\label{Ecscaledlinear}
\end{eqnarray} 
leading to the linearly scaled one-parameter double-hybrid (LS1DH) approximation
\begin{eqnarray}
E^{\text{LS1DH},\l}_{xc} &=& \l E_x^{\HF} + (1-\l) E_x[n] 
\nonumber\\
&&+ (1-\l^3) E_c[n] + \l^3 E_c^{\text{MP2}},
\label{LS1DH}
\end{eqnarray} 
which works reasonably well for PBE and corresponds to the form used for the PBE0-DH~\cite{BreAda-JCP-11} and PBE0-2~\cite{ChaMao-CPL-12} double hybrids. Finally, note that Fromager and coworkers~\cite{Fro-JCP-11,CorFraTeaFro-MP-13} explored the theoretical basis of the two-parameter double-hybrid approximations and proposed new double-hybrid schemes.

The vast majority of all these double hybrids have been applied with generalized-gradient approximation (GGA) density functionals. Only recently, Goerigk and Grimme~\cite{GoeGri-JCTC-11} proposed two-parameter double-hybrid schemes based on meta-GGA density functionals. In these schemes, the opposite-spin MP2 correlation energy is combined with a refitted Tao-Perdew-Staroverov-Scuseria (TPSS)~\cite{TaoPerStaScu-PRL-03} meta-GGA exchange-correlation functional, or with a refitted Perdew-Wang~\cite{Per-INC-91} GGA exchange functional and a refitted Becke-95 (B95)~\cite{Bec-JCP-96} meta-GGA correlation functional. Kozuch and Martin~\cite{KozMar-PCCP-11,KozMar-JCC-13} also tested numerous spin-component-scaled double hybrids, with four or more optimized parameters, including the B95 meta-GGA correlation functional and the TPSS, B98~\cite{Bec-JCP-98}, BMK~\cite{BoeMar-JCP-04}, $\tau$HCTH~\cite{BoeHan-JCP-02} meta-GGA exchange-correlation functionals. 

In this work, we reexamine the theoretical basis of double-hybrid approximations using meta-GGA functionals, and we construct one-parameter double hybrids using the TPSS meta-GGA functional. While the 1DH and LS1DH schemes can be readily applied with a meta-GGA functional, the DS1DH scheme requires an extension of the density scaling to the non-interacting kinetic energy density $\tau(\b{r})$ that is used in the TPSS functional. We then assess the accuracy of these one-parameter meta-GGA double hybrids on test sets of atomization energies and reaction barrier heights.

\section{Double-hybrid meta-GGA approximations}

In addition to the explicit dependence on the density $n(\b{r})$, its gradient $\bm{\nabla} n(\b{r})$ and possibly its Laplacian $\bm{\nabla}^2 n(\b{r})$, a meta-GGA density functional also generally depends on the non-interacting positive kinetic energy density $\tau(\b{r})$ which can be seen as an implicit functional of the density (for simplicity, we only write the equations for the spin-unpolarized case; the extension to the general spin-polarized case is straightforward):
\begin{eqnarray}
\tau[n](\b{r}) = \bra{\Phi[n]} \hat{\tau}(\b{r}) \ket{\Phi[n]} = \frac{1}{2} \sum_i^{\text{occ}} \left| \bm{\nabla}_{\b{r}} \phi_i[n] (\b{r})\right|^2,
\end{eqnarray} 
where the positive kinetic energy density operator $\hat{\tau}(\b{r}) = (1/2) \sum_{\sigma=\uparrow,\downarrow}\bm{\nabla}_{\b{r}}\, \hat{\psi}_{\sigma}^\dagger(\b{r}) \cdot \bm{\nabla}_{\b{r}} \, \hat{\psi}_{\sigma}(\b{r})$ has been introduced (with the creation and annihilation field operators $\hat{\psi}_{\sigma}^\dagger(\b{r})$ and $\hat{\psi}_{\sigma}(\b{r})$), $\Phi[n]$ is the (KS) single-determinant wave function minimizing the kinetic energy $\bra{\Phi} \hat{T} \ket{\Phi}$ and giving the density $n$, and $\phi_i[n] (\b{r})$ are the associated (KS) spin-orbitals.

Defining the scaled non-interacting kinetic energy density $\tau_\g$ as the non-interacting kinetic energy density corresponding the scaled density $n_\g(\b{r})=\g^3 n(\g\b{r})$ (where $\g$ is an arbitrary positive scaling factor), one can show
\begin{eqnarray}
\tau_\g[n](\b{r}) &\equiv& \tau[n_{\g}](\b{r}) = \frac{1}{2} \sum_i^{\text{occ}} \left| \bm{\nabla}_{\b{r}} \phi_i[n_\g] (\b{r})\right|^2
\nonumber\\
                    &=& \frac{\g^3}{2} \sum_i^{\text{occ}} \left| \bm{\nabla}_{\b{r}} \phi_i[n] (\g \b{r})\right|^2
\nonumber\\
                    &=& \frac{\g^5}{2} \sum_i^{\text{occ}} \left| \bm{\nabla}_{\g \b{r}} \phi_i[n] (\g \b{r})\right|^2
\nonumber\\
                    &=& \g^5 \tau[n](\g \b{r}),
\label{taug}
\end{eqnarray}
where the scaling relation $\phi_i[n_\g] (\b{r}) = \g^{3/2} \phi_i[n] (\g \b{r})$ has been used. The scaling relation on $\tau[n](\b{r})$ is consistent with the well-known quadratic scaling of the non-interacting kinetic energy $T_s[n_\g]=\g^2 T_s[n]$ where $T_s[n]=\int \tau[n](\b{r}) d\b{r}$~\cite{Sha-PRA-70}. Furthermore, this scaling relation can easily be verified for one-electron systems with the von Weizs\"acker kinetic energy density, $\tau^W(\b{r}) = (1/8) \left| \bm{\nabla}_{\b{r}} n(\b{r}) \right|^2 / n(\b{r})$, and for the uniform electron gas, $\tau^{\text{unif}} = (3/10) (3\pi^2)^{2/3} n^{5/3}$.

Working with $\tau[n](\b{r})$ as an implicit functional of the density in a self-consistent KS calculation requires to use the optimized effective potential approach to calculate the functional derivative of the exchange-correlation energy $E_{xc}[n]$ with respect to the density $n$~\cite{ArbKau-CPL-03}, which is computationally impractical. The standard practice~\cite{NeuNobHan-MP-96,NeuHan-CPL-96,AdaErnScu-JCP-00,ArbKauMalRevMal-PCCP-02,FurPer-JCP-06,SunMarCsoRuzHaoKimKrePer-PRB-11,ZahLeaGor-JCP-13} is to consider the exchange-correlation energy as an explicit functional of both $n$ and $\tau$, $E_{xc}[n,\tau]$. Following this practice, we define a density-scaled one-parameter hybrid (DS1H) approximation with a $\tau$-dependent functional as
\begin{eqnarray}
E^{\text{DS1H},\l} &=& \min_{\Phi} \Bigl\{\bra{\Phi}\hat{T}+\hat{V}_{\text{ext}}+\l\hat{W}_{ee} \ket{\Phi}
\nonumber\\
&&+\bar{E}_{\H}^{\l}[n_{\Phi}]+\bar{E}_{xc}^{\l}[n_{\Phi},\tau_{\Phi}]\Bigl\}, 
\label{DS1H}
\end{eqnarray} 
where $\Phi$ is a single-determinant wave function, $\hat{T}$ is the kinetic energy operator, $\hat{V}_{\text{ext}}$ is the external (e.g., electron-nucleus) potential operator, $\hat{W}_{ee}$ is the Coulomb electron-electron interaction operator, $\bar{E}_{\H}^{\l}[n_{\Phi}]$ and $\bar{E}_{xc}^{\l}[n_{\Phi},\tau_{\Phi}]$ are the complement Hartree and exchange-correlation functionals evaluated at the density and kinetic energy density of $\Phi$, $n_{\Phi}(\b{r}) = \bra{\Phi} \hat{n}(\b{r}) \ket{\Phi}$ and $\tau_{\Phi}(\b{r}) = \bra{\Phi} \hat{\tau}(\b{r}) \ket{\Phi}$. The complement Hartree and exchange functionals are linear with respect to $\l$:
\begin{eqnarray}
\bar{E}_{\H}^{\l}[n]=(1-\l) E_{\H}[n],
\end{eqnarray} 
\begin{eqnarray}
\bar{E}_{x}^{\l}[n,\tau]=(1-\l) E_{x}[n,\tau],
\end{eqnarray} 
where $E_{\H}[n]$ and $E_{x}[n,\tau]$ are the usual KS Hartree and exchange functionals. The complement correlation functional is obtained via the extension of uniform coordinate scaling of the density~\cite{LevPer-PRA-85,LevYanPar-JCP-85,Lev-PRA-91,LevPer-PRB-93} to the kinetic energy density
\begin{eqnarray}
\bar{E}_{c}^{\l}[n,\tau] &=& E_{c}[n,\tau]-E_{c}^{\l}[n,\tau]
\nonumber\\
                         &=& E_{c}[n,\tau]-\l^2 E_{c}[n_{1/\l},\tau_{1/\l}],
\label{}
\end{eqnarray} 
where $E_{c}[n,\tau]$ is the usual KS correlation functional, $E_{c}^{\l}[n,\tau]$ is the correlation functional corresponding to the interaction $\l\hat{W}_{ee}$, $n_{1/\l}(\b{r})=(1/\l)^3 n(\b{r}/\l)$ is the scaled density and $\tau_{1/\l}(\b{r})=(1/\l)^5 \tau(\b{r}/\l)$ is the scaled kinetic energy density.

The minimizing single-determinant wave function $\Phi^\l$ in Eq.~(\ref{DS1H}) is calculated by the self-consistent eigenvalue equation:
\begin{eqnarray}
\left( \hat{T}+\hat{V}_{\text{ext}}+\l \hat{V}_{\H x}^{\HF}[\Phi^\l] + \hat{V}_{\H}^{\l}[n_{\Phi^\l}] + \hat{V}_{xc}^{\l}[n_{\Phi^\l},\tau_{\Phi^\l}] \right) \ket{\Phi^\l}
\nonumber\\
= {\cal E}_0^{\l} \ket{\Phi^\l}, \:\;\;\;\;\;\;\;
\label{DS1Heigenval}
\end{eqnarray} 
where $\hat{V}_{\H x}^{\HF}$ is the nonlocal HF potential operator, $\hat{V}_{\H}^{\l}$ is the complement local Hartree potential operator, and $\hat{V}_{xc}^{\l}$ is the complement exchange-correlation potential operator
\begin{eqnarray}
\hat{V}_{xc}^{\l}[n,\tau] = \int \frac{\delta \bar{E}_{xc}^\l[n,\tau]}{\delta n(\b{r})} \hat{n}(\b{r}) d\b{r} + \int \frac{\delta \bar{E}_{xc}^\l[n,\tau]}{\delta \tau(\b{r})} \hat{\tau}(\b{r}) d\b{r},
\nonumber\\
\label{Vxclnt}
\end{eqnarray} 
where $\hat{n}(\b{r})=\sum_{\sigma=\uparrow,\downarrow} \hat{\psi}_{\sigma}^\dagger(\b{r}) \hat{\psi}_{\sigma}(\b{r})$ is the density operator and the second term in Eq.~(\ref{Vxclnt}) corresponds to a non-multiplicative ``potential'' operator.
%%%%%%%%%%%%%%%%%%%%%%%%%%%%%%%%%%%%%%%%%%%%%%%%%%%%%%%%%%%%%%%%%%%%%%%%%%%%%%%%%%%%%%%%%%%%
\begin{figure*}
\includegraphics[scale=0.30,angle=-90]{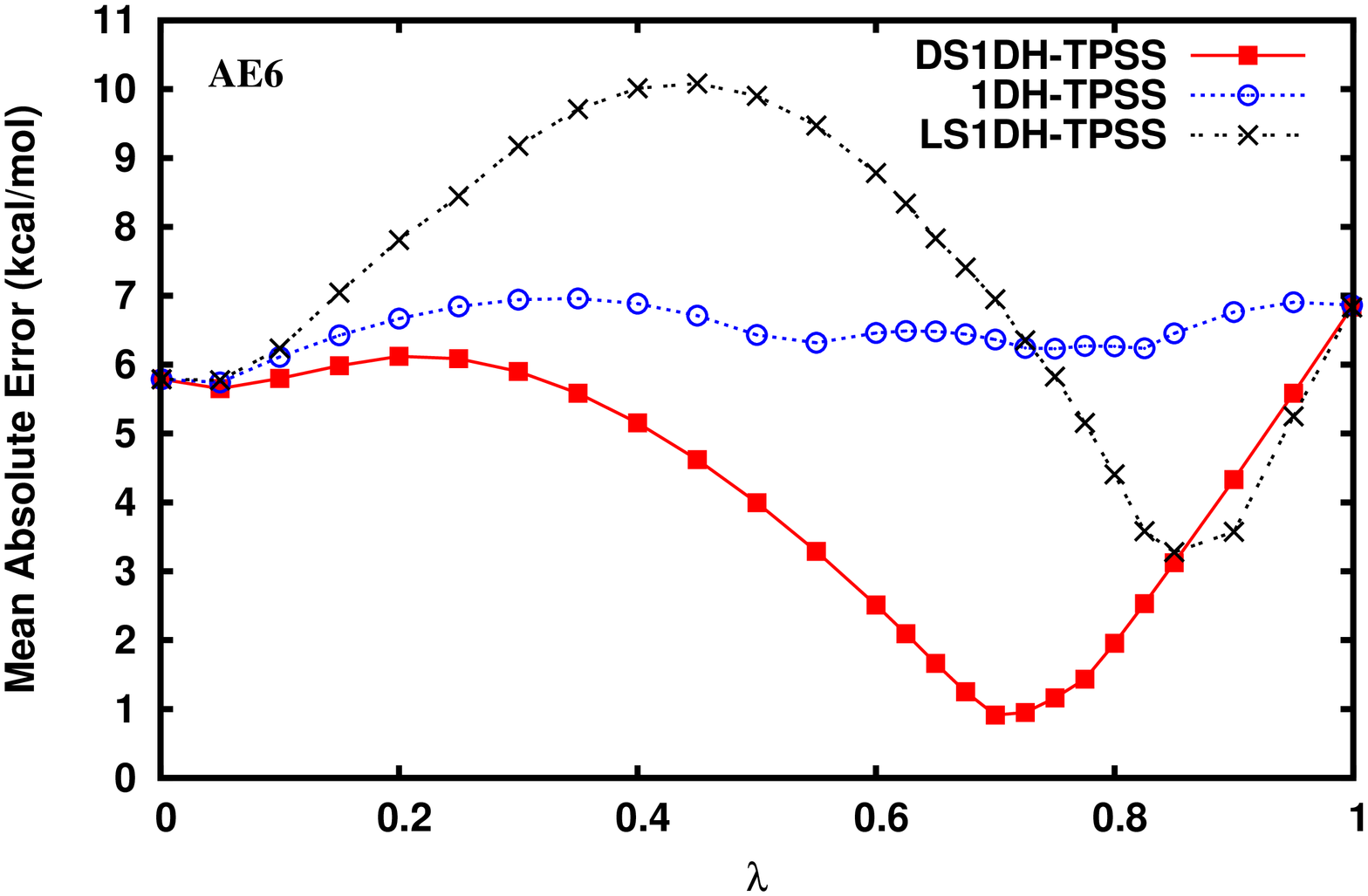}
\includegraphics[scale=0.30,angle=-90]{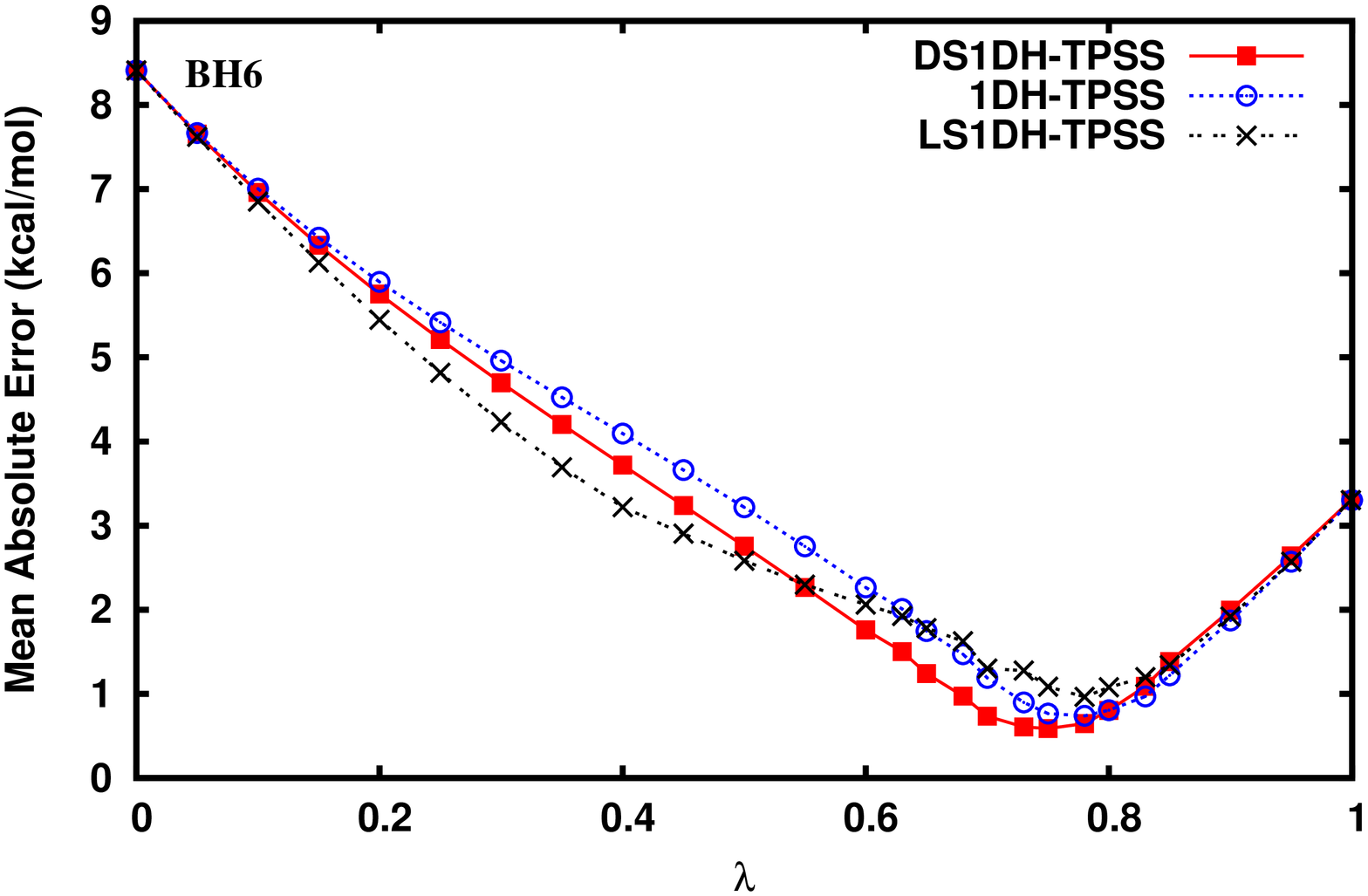}
\caption{MAEs for the AE6 (left) and BH6 (right) test sets as functions of the parameter $\l$ for the DS1DH, 1DH and LS1DH approximations with the TPSS exchange-correlation density functional. All calculations were carried out with the cc-pVQZ basis set.
}
\label{fig:AE6BH6}
\end{figure*}
%%%%%%%%%%%%%%%%%%%%%%%%%%%%%%%%%%%%%%%%%%%%%%%%%%%%%%%%%%%%%%%%%%%%%%%%%%%%%%%%%%%%%%%%%%%%%

The nonlinear Rayleigh-Schr\"odinger perturbation theory of Refs.~\onlinecite{AngGerSavTou-PRA-05,FroJen-PRA-08,Ang-PRA-08,ShaTouSav-JCP-11} can readily be extended to start with the DS1H reference of Eq.~(\ref{DS1H}) with a $\tau$-dependent functional (details are given in the supplementary material~\cite{SouShaTou-JJJ-XX-sup}). For the second-order energy correction, due to Brillouin's theorem, only double excitations contribute and consequently the nonlinear terms of the perturbation theory vanish since they involve expectation values of the one-electron operators $\hat{n}(\b{r})$ and $\hat{\tau}(\b{r})$ between determinants differing by two spin orbitals. The second-order energy correction to be added to the DS1H energy has thus a standard MP2 form
\begin{eqnarray}
E^{\l,(2)} &=& \l^2 \sum_{i<j\atop a<b} \frac{\left|\bra{ij}\ket{ab}\right|^2}{\varepsilon_{i} +\varepsilon_{j} -\varepsilon_{a} -\varepsilon_{b}} = \l^2 E_c^{\text{MP2}},
\label{MP2}
\end{eqnarray}
where $i,j$ and $a,b$ refer to occupied and virtual DS1H spin-orbitals, respectively, with associated orbital eigenvalues $\varepsilon_{k}$, and $\bra{ij}\ket{ab}$ are the antisymmetrized two-electron integrals. The DS1DH exchange-correlation energy for a $\tau$-dependent functional is thus (dropping from now on the explicit dependence on $\Phi^{\l}$)
\begin{eqnarray}
E^{\text{DS1DH},\l}_{xc} &=& \l E_x^{\HF} + (1-\l) E_x[n,\tau] 
\nonumber\\
&&+ E_c[n,\tau] -\l^2 E_c[n_{1/\l},\tau_{1/\lambda}] + \l^2 E_c^{\text{MP2}}.
\nonumber\\
\label{ExcDS1DHtau}
\end{eqnarray} 
Neglecting the scaling in the correlation functional, $E_{c}[n_{1/\l},\tau_{1/\l}] \approx E_{c}[n,\tau]$, gives the 1DH exchange-correlation energy
\begin{eqnarray}
E^{\text{1DH},\l}_{xc} &=& \l E_x^{\HF} + (1-\l) E_x[n,\tau] 
\nonumber\\
&&+ (1-\l^2) E_c[n,\tau] + \l^2 E_c^{\text{MP2}},
\label{Exc1DHtau}
\end{eqnarray} 
and using the approximate scaling $E_c[n_{1/\l},\tau_{1/\l}] \approx (1-\l) E_c^{\text{MP2}} + \l E_c[n,\tau]$ gives the LS1DH exchange-correlation energy
\begin{eqnarray}
E^{\text{LS1DH},\l}_{xc} &=& \l E_x^{\HF} + (1-\l) E_x[n,\tau] 
\nonumber\\
&&+ (1-\l^3) E_c[n,\tau] + \l^3 E_c^{\text{MP2}}.
\label{LS1DHtau}
\end{eqnarray} 
We apply these double-hybrid schemes with the TPSS exchange-correlation functional and refer to them as DS1DH-TPSS, 1DH-TPSS and LS1DH-TPSS.

%%%%%%%%%%%%%%%%%%%%%%%%%%%%%%%%%%%%%%%%%%%%%%%%%%%%%%%%%%%%%%%%%%%%%%%%%%%%%%%%%%%%%%%%%%%%%%%%%%%%%%%%%%%%%%%%%%%%%%%%%%%%%%%%%%%%
\begingroup
\squeezetable
\begin{table*}[t]
\caption{MAEs and MEs (in kcal/mol) on the AE6 and BH6 test sets for several methods. For the double-hybrid DS1DH, 1DH, and LS1DH approximations, the results are for the optimal values of $\l$ which minimize the MAEs of the AE6 and BH6 sets, separately. All calculations were carried out with the cc-pVQZ basis set.}
\label{tab:AE6BH6}
\begin{tabular}{llrrcllrr}
\hline
\hline
           & \multicolumn{3}{c}{AE6}          &                &  \multicolumn{3}{c}{BH6}  \\
                        \cline{2-4}                                    \cline{6-8}
Method     & $\l$              &MAE  &  ME    & \phantom{xxxx} &  $\l$      & MAE  &  ME  \\
\hline                                                   
BLYP$^a$   &                   & 6.52 & -1.18 &                &            & 8.10 &-8.10 \\
PBE$^a$    &                   & 15.5 & 12.4  &                &            & 9.61 &-9.61 \\
TPSS       &                   & 5.79 & 3.78  &                &            & 8.41 &-8.41 \\
MP2$^a$    &                   & 6.86 & 4.17  &                &            & 3.32 & 3.11 \\
\\
DS1DH-BLYP$^a$ & $\l=0.80$     & 4.73 & -2.52 &                &  $\l=0.65$ & 0.60 & 0.24 \\
1DH-BLYP$^a$   & $\l=0.55$     & 1.46 & 0.07  &                &  $\l=0.75$ & 0.80 &-0.18 \\
B2-PLYP$^a$    &               & 1.39 & -1.09 &                &            & 2.21 &-2.21 \\[0.1cm]

DS1DH-PBE$^a$  & $\l=0.65$     & 3.78 & 1.30  &                &  $\l=0.80$ & 1.32 & 0.48\\
1DH-PBE$^a$    & $\l=0.55^b$   & 8.64$^b$ & 7.06$^b$  &        &  $\l=0.80$ & 1.42 & 0.12 \\
LS1DH-PBE$^c$  & $\l=0.75$     & 3.59 & 0.23  &                &  $\l=0.70$ & 0.73 & -0.20 \\[0.1cm]

DS1DH-TPSS & $\l=0.70$         & 0.91 & 0.07  &                &  $\l=0.75$ & 0.59 & -0.14 \\
1DH-TPSS   & $\l=0.05$         & 5.74 & 2.37  &                &  $\l=0.78$ & 0.74 & -0.28 \\
LS1DH-TPSS & $\l=0.85$         & 3.28 & 2.21  &                &  $\l=0.78$ & 0.96 &  0.13 \\

\hline
\hline
\multicolumn{9}{l}{$^a$ Data from Ref.~\onlinecite{ShaTouSav-JCP-11}.}\\
\multicolumn{9}{l}{$^b$ This is a local minimum. The global minimum is for $\l=1.0$, i.e. MP2.}\\
\multicolumn{9}{l}{$^c$ Data from Ref.~\onlinecite{TouShaBreAda-JCP-11}.}
\end{tabular}
\end{table*}
\endgroup
%%%%%%%%%%%%%%%%%%%%%%%%%%%%%%%%%%%%%%%%%%%%%%%%%%%%%%%%%%%%%%%%%%%%%%%%%%%%%%%%%%%%%%%%%%%%%%%%%%%%%%%%%%%%%%%%%%%%%%%%%%%%%%%%%%%%

%%%%%%%%%%%%%%%%%%%%%%%%%%%%%%%%%%%%%%%%%%%%%%%%%%%%%%%%%%%%%%%%%%%%%%%%%%%%%%%%%%%%%%%%%%%%%%%%%%%%%%%%%%%%%%%%%%%%%%%%%%%%%%%%%%%%
\begingroup
\squeezetable
\begin{table*}[t]
\caption{Atomization energies (in kcal/mol) of the 49 molecules of the set of Ref.~\onlinecite{FasCorSanTru-JPCA-99} (G2-1 test set except for the six molecules containing Li, Be, and Na). The calculated values were obtained using the functional TPSS and the double hybrids DS1DH-TPSS (with $\l=0.725$) and B2-PLYP with the cc-pVQZ basis set and MP2(full)/6-31G* geometries. The zero-point energies are removed in the reference values. For each method, the value with the largest error is indicated in boldface.}
\label{tab:G55}
\begin{tabular}{lcccc}
\hline
\hline

Molecule                 & \multicolumn{1}{c}{TPSS} & \multicolumn{1}{c}{DS1DH-TPSS} & \multicolumn{1}{c}{B2-PLYP$^a$} & \multicolumn{1}{c}{Reference$^b$} \\
\hline
CH	                 &  86.16              &   81.79              &        83.70  &    84.00 \\
CH$_2$ ($^{3}$B$_{1}$)	 &  197.58             &   192.21             &       190.57  &    190.07\\
CH$_2$ ($^{1}$A$_{1}$)   &  180.14             &   176.42             &       178.84  &    181.51\\
CH$_3$	                 &  313.08             &   307.41             &       307.90  &    307.65\\
CH$_4$	                 &  424.24             &   418.78             &       419.19  &    420.11\\
NH	                 &  90.06              &   81.87              &        84.89  &    83.67 \\
NH$_2$	                 &  187.57             &   179.47             &       183.94  &   181.90\\
NH$_3$	                 &  299.39             &   293.93             &       297.69  &   297.90 \\
OH	                 &  106.17             &   105.07             &       106.43  &   106.60\\
OH$_2$	                 &  227.61             &   229.70             &       229.81  &   232.55\\
FH	                 &  137.73             &   139.98             &       139.00  &   141.05\\
SiH$_2$ ($^{1}$A$_{1}$)	 &  155.34             &   149.73             &       151.77  &   151.79  \\
SiH$_2$ ($^{3}$B$_{1}$)	 &  140.29             &   133.77             &       131.78  &  131.05\\
SiH$_3$                  &  235.86             &   227.23             &       226.67  &  227.37 \\
SiH$_4$                  &  332.36             &   322.33             &       321.95  & 322.40  \\
PH$_2$	                 &  160.61             &   150.92             &       154.80  & 153.20  \\
PH$_3$                   &  248.32             &   237.11             &       240.73  & 242.55 \\
SH$_2$                   &  183.95             &   180.31             &       180.58  &  182.74 \\
ClH	                 &  105.42             &   105.72             &       105.01  &  106.50 \\
HCCH	                 &  403.17             &   404.25             &      404.45   & 405.39  \\
H$_2$CCH$_2$	         &  566.27             &   561.66             &      562.15   & 563.47 \\
H$_3$CCH$_3$             &  717.01             &   710.97             &      710.22   &  712.80 \\
CN	                 &  180.61             &   {\bf 173.89}       &      179.61   & 180.58 \\
HCN	                 &  312.39             &   311.53             &      314.12   & 313.20 \\
CO	                 &  253.19             &   258.75             &      258.28   & 259.31 \\
HCO	                 &  281.80             &   279.94             &      280.62   & 278.39 \\
H$_2$CO	                 &  375.03             &   373.44             &      373.56   & 373.73 \\
H$_3$COH	         &  513.26             &   510.78             &     510.38    & 512.90 \\
N$_2$	                 &  226.73             &   224.84             & 229.24	    &228.46 \\
H$_2$NNH$_2$             &  443.05             &   432.05             & 438.77	    &  438.60\\
NO	                 &  156.29             &   151.51             &  155.04       &  155.22 \\
O$_2$	                 &  126.48             &   121.49             & 122.71        & 119.99  \\
HOOH	                 &  268.05             &   264.63             &  265.44       & 268.57 \\
F$_2$	                 &  45.36              &   36.46              & 36.29	    & 38.20  \\
CO$_2$	                 &  388.56             &   392.98             &  391.23       & 389.14 \\
Si$_2$	                 &  75.69              &   72.62              & 70.58	    & 71.99 \\
P$_2$	                 &  114.84             &   110.56             &   115.84      & 117.09 \\
S$_2$	                 &  106.79             &   101.91             &  102.27       & 101.67 \\
Cl$_2$                   &  58.36              &   56.97              & 55.48	    & 57.97 \\
SiO	                 &  184.87             &   190.16             & 190.82	    &  192.08 \\
SC	                 &  167.49             &   169.02             &  168.86       &  171.31 \\
SO	                 &  127.99             &   124.31             & 125.33	    & 125.00 \\
ClO	                 &  69.45              &   59.78              &   62.70       &  64.49 \\
ClF	                 &  64.48              &   60.38              &  59.85	    & 61.36 \\
Si$_2$H$_6$              &  {\bf 545.77}       &   531.14             &  529.02       & 530.81 \\
CH$_3$Cl	         &  396.93             &   394.41             & 392.62	    & 394.64 \\
CH$_3$SH	         &  476.80             &   471.71             & 470.71	    & 473.84 \\
HOCl	                 &  164.67             &   162.66             & 162.27	    & 164.36 \\
SO$_2$	                 &  252.11             &   254.27             & {\bf 251.10}  &257.86 \\
\hline                                                                    
MAE                      &  3.9                &    2.3      &  1.6          &      \\         
ME                       &  2.2                &   -1.7      &  -1.0         &      \\ 
\hline
\hline
$^a$Data from Ref.~\onlinecite{ShaTouSav-JCP-11}.\\
$^b$From Ref.~\onlinecite{FasCorSanTru-JPCA-99}.\\
\end{tabular}
\end{table*}
\endgroup
%%%%%%%%%%%%%%%%%%%%%%%%%%%%%%%%%%%%%%%%%%%%%%%%%%%%%%%%%%%%%%%%%%%%%%%%%%%%%%%%%%%%%%%%%%%%%%%%%%%%%%%%%%%%%%%%%%%%%%%%%%%%%%%%%%%%

\section{Computational details}

Calculations have been performed with a development version of the MOLPRO program~\cite{Molproshort-PROG-12}, in which the DS1DH-TPSS, 1DH-TPSS and LS1DH-TPSS approximations have been implemented. The scaling relations for the scaled exchange-correlation energy and its derivatives for a general meta-GGA functional are given in the appendix. The empirical parameter $\l$ is optimized on the AE6 and BH6 test sets~\cite{LynTru-JPCA-03}. The AE6 set is a small representative benchmark set of six atomization energies consisting of SiH$_4$, S$_2$, SiO, C$_3$H$_4$ (propyne), C$_2$H$_2$O$_2$ (glyoxal), and C$_4$H$_8$ (cyclobutane). The BH6 set is a small representative benchmark set of forward and reverse hydrogen barrier heights of three reactions, OH + CH$_4$ $\to$ CH$_3$ + H$_2$O, H + OH $\to$ O + H$_2$, and H + H$_2$S $\to$ HS + H$_2$. All the calculations for the AE6 and BH6 sets were performed at the optimized QCISD/MG3 geometries~\cite{LynZhaTru-JJJ-XX} using the Dunning cc-pVQZ basis set~\cite{Dun-JCP-89,WooDun-JCP-93}. The performance of the best double hybrid is then checked on the larger benchmark set of 49 atomization energies of Ref.~\onlinecite{FasCorSanTru-JPCA-99} (G2-1 test set~\cite{CurRagTruPop-JCP-91,CurRagRedPop-JCP-97} except for the six molecules containing Li, Be, and Na) at MP2(full)/6-31G* geometries using the Dunning cc-pVQZ basis set. Core electrons are kept frozen in all our MP2 calculations. Spin-restricted calculations are performed for all the closed-shell systems, and spin-unrestricted calculations for all the open-shell systems.

\section{Results and discussion}

In Figure~\ref{fig:AE6BH6}, we plot the mean absolute errors (MAEs) for the AE6 and BH6 test sets as functions of the parameter $\l$ for the DS1DH-TPSS, 1DH-TPSS and LS1DH-TPSS approximations. The MAEs and mean errors (MEs) of the double hybrids based on the TPSS functional at the optimal values of $\l$ which minimize the MAEs on the AE6 and BH6 sets are also reported in Table~\ref{tab:AE6BH6}, and compared to those obtained with standard BLYP, PBE, TPSS and MP2, as well as with other double-hybrid approximations based on the BLYP and PBE functionals.

For $\l=0$, all these double hybrids reduce to a standard KS calculation with the TPSS functional, while for $\l=1$ they all reduce to a standard MP2 calculation. For the AE6 set, DS1DH-TPSS gives by far the smallest MAE with 0.91 kcal/mol at the optimal value of $\l=0.70$. LS1DH-TPSS gives a larger MAE of 3.28 kcal/mol for an optimal value of $\l=0.85$, and 1DH-TPSS gives a yet larger MAE of 5.74 kcal/mol for an optimal value of $\l=0.05$, providing virtually no improvement over the TPSS KS calculation at $\l=0$. For the BH6 set, the three double-hybrid approximations are very similar for the entire range of $\lambda$, indicating that the scaling in the correlation functional contribution is not as crucial for barrier heights as for atomization energies. The MAE minima are 0.59 kcal/mol at $\lambda=0.75$, 0.74 kcal/mol at $\lambda=0.78$ and 0.96 kcal/mol at $\lambda=0.78$ for DS1DH-TPSS, 1DH-TPSS, and LS1DH-TPSS, respectively.

Neglecting the scaling of the density and of the kinetic energy density in the TPSS correlation functional, $E_{c}[n_{1/\l},\tau_{1/\l}] \approx E_{c}[n,\tau]$, i.e. going from DS1DH-TPSS to 1DH-TPSS, largely deteriorates the accuracy of atomization energies. A similar deterioration is obtained when neglecting the scaling of the density in the PBE correlation functional, whereas a large improvement of atomization energies is observed when neglecting the scaling of the density in the LYP correlation functional (see Table~\ref{tab:AE6BH6} and Ref.~\onlinecite{ShaTouSav-JCP-11}). These different behaviors between the PBE and TPSS correlation functionals on a one hand and the LYP correlation functional of the second hand may be related to the observation on atoms with non-degenerate KS systems that PBE and TPSS are more accurate than LYP for the high-density limit (or weak-interaction limit), $E_c^{(2)} = \lim_{\l\to0}E_{c}[n_{1/\l}]$~\cite{IvaLev-JPCA-98,StaScuPerTaoDav-PRA-04,WhiBur-JCP-05,StaScuPerDavKat-PRA-06}.

Contrary to most other double hybrids, the DS1DH-TPSS double-hybrid approximation gives very close optimal values of $\l$ on the AE6 and BH6 sets, i.e. $\lambda=0.70$ and $\lambda=0.75$. For general applications, we propose to use the value of $\lambda=0.725$ for this double hybrid when using the cc-pVQZ basis set.

Finally, in Table~\ref{tab:G55}, we compare DS1DH-TPSS (with the optimal parameter $\l=0.725$) with TPSS and the standard double hybrid B2-PLYP~\cite{Gri-JCP-06} on the larger set of 49 atomization energies of Ref.~\onlinecite{FasCorSanTru-JPCA-99}. With MAEs of 3.9 and 2.3 kcal/mol for TPSS and DS1DH-TPSS, respectively, it is clear that the improvement in accuracy brought by DS1DH-TPSS over TPSS observed on the small AE6 set remains (although smaller) for this larger set. DS1DH-TPSS is however slightly less accurate on average on this set than B2-PLYP (MAE of 1.6 kcal/mol).

\section{Conclusions}

We have constructed one-parameter double-hybrid approximations using the TPSS meta-GGA exchange-correlation functional and tested them on test sets of atomization energies and reaction barrier heights. We have shown that neglecting the scaling of the density and of the kinetic energy density in the correlation functional largely deteriorates the accuracy on atomization energies, in contrast to what was previously found for double hybrids based on the BLYP functional. We thus propose the density-scaled double-hybrid DS1DH-TPSS approximation with a fraction of HF exchange of $\l=0.725$ as a viable meta-GGA double hybrid for thermochemistry calculations, improving over both standard KS TPSS and MP2 calculations. We hope that this work will lead to more investigations of meta-GGA double hybrids with minimal empiricism. Possible extensions of this work include introducing meta-GGA functionals in multiconfigurational hybrids~\cite{ShaSavJenTou-JCP-12} or in Coulomb-attenuated double hybrids~\cite{CorFro-ARX-13}.

\section*{Acknowledgments}
We thank Andreas Savin (UPMC/CNRS, Paris) for stimulating discussions.

\appendix
\section{Scaling relations for a meta-GGA correlation energy functional and its derivatives}
\label{app:scaling}

We give the expressions for the scaled meta-GGA correlation functional $E_c^\l[n,\tau]=\l^2 E_c[n_{1/\l},\tau_{1/\l}]$ and its derivatives (see, also, Refs.~\onlinecite{ShaTouSav-JCP-11,ShaSavJenTou-JCP-12}). Starting from a standard meta-GGA density functional, depending on the density $n(\b{r})$, the square of the density gradient $\left|\bm{\nabla}_\b{r} n(\b{r})\right|^2$, the Laplacian of the density $\bm{\nabla}_\b{r}^2 n(\b{r})$ and/or the non-interacting kinetic energy density $\tau(\b{r})$
\begin{eqnarray}
E_{c}[n,\tau] &=& \int e_{c} \left(n(\b{r}),\left|\bm{\nabla}_\b{r} n(\b{r})\right|^2, \bm{\nabla}_\b{r}^2 n(\b{r}), \tau(\b{r}) \right) d\b{r},
\nonumber\\
\end{eqnarray}
the corresponding scaled functional is written as
\begin{eqnarray}
E_{c}^\l[n,\tau] &=& \int e_{c}^\l \left(n(\b{r}),\left|\bm{\nabla}_\b{r} n(\b{r})\right|^2, \bm{\nabla}_\b{r}^2 n(\b{r}), \tau(\b{r}) \right) d\b{r},
\nonumber\\
\end{eqnarray}
where the energy density is obtained by the scaling relation
\begin{eqnarray}
e_{c}^\l = \l^5 e_{c} \left(\frac{n(\b{r})}{\l^3},\frac{\left|\bm{\nabla}_\b{r} n(\b{r})\right|^2}{\l^8}, \frac{\bm{\nabla}_\b{r}^2 n(\b{r})}{\l^5}, \frac{\tau(\b{r})}{\l^5}\right).
\end{eqnarray}
The first-order derivatives of the energy density are
\begin{eqnarray}
\frac{\partial e_c^{\l}}{\partial n} = \l^2 \frac{\partial e_c}{\partial n} \left(\frac{n(\b{r})}{\l^3},\frac{\left|\bm{\nabla}_\b{r} n(\b{r})\right|^2}{\l^8} ,\frac{\bm{\nabla}_\b{r}^2 n(\b{r})}{\l^5}, \frac{\tau(\b{r})}{\l^5}\right),
\end{eqnarray}
and
\begin{eqnarray}
\frac{\partial e_c^{\l}}{\partial \left|\bm{\nabla} n \right|^2} = \frac{1}{\l^3} \frac{\partial e_c}{\partial \left|\bm{\nabla} n \right|^2} \left(\frac{n(\b{r})}{\l^3},\frac{\left|\bm{\nabla}_\b{r} n(\b{r})\right|^2}{\l^8} ,\frac{\bm{\nabla}_\b{r}^2 n(\b{r})}{\l^5}, \frac{\tau(\b{r})}{\l^5}\right),
\nonumber\\
\end{eqnarray}
and
\begin{eqnarray}
\frac{\partial e_c^{\l}}{\partial \left|\bm{\nabla}^2 n \right|} = \frac{\partial e_c}{\partial \left|\bm{\nabla}^2 n \right|} \left(\frac{n(\b{r})}{\l^3},\frac{\left|\bm{\nabla}_\b{r} n(\b{r})\right|^2}{\l^8} ,\frac{\bm{\nabla}_\b{r}^2 n(\b{r})}{\l^5}, \frac{\tau(\b{r})}{\l^5}\right),
\nonumber\\
\end{eqnarray}
and
\begin{eqnarray}
\frac{\partial e_c^{\l}}{\partial \tau} = \frac{\partial e_c}{\partial \tau} \left(\frac{n(\b{r})}{\l^3},\frac{\left|\bm{\nabla}_\b{r} n(\b{r})\right|^2}{\l^8} ,\frac{\bm{\nabla}_\b{r}^2 n(\b{r})}{\l^5}, \frac{\tau(\b{r})}{\l^5}\right).
\end{eqnarray}
The same scaling relations apply for spin-dependent functionals $E_{c}[n_{\uparrow},n_{\downarrow},\tau_{\uparrow},\tau_{\downarrow}]$.

% BIBLIOGRAPHY---------------------------------------------

\end{document}